\documentclass[aps, pra, reprint, groupedaddress, superscriptaddress, footinbib, floatfix]{revtex4-1}

\usepackage{amsmath, amssymb, bm}
\usepackage{graphicx}
\usepackage{natbib}
\usepackage{float}
\usepackage[euler]{textgreek}
\usepackage[english]{babel} % Changes captions and titles if english not chosen :/ See \bblindtext command below
\usepackage{blindtext}
\usepackage[dvipsnames]{xcolor}
\usepackage{braket}
\usepackage{hyperref}
\definecolor{ddblue}{RGB}{0,0,160}
\hypersetup{colorlinks=true,linktoc=all,linkcolor=ddblue,citecolor=blue,urlcolor=[rgb]{0, 0.38, 0}}

\newcommand*\vecc[1]{\ifmmode\bm{#1}\else\textbf{#1}\fi}

\begin{document}
	
	\preprint{APS/123-QED}
	
	%title{Interference of Raman transitions driven by multiple phase-locked optical frequencies in a cavity atom interferometer}
	%\title{Properties of Raman transitions driven by multiple phase-locked optical frequency components in an optical cavity}

	%\title{Interference of Raman transitions driven by phase-modulated light in a cavity atom interferometer}
	\title{Raman transitions driven by phase-modulated light in a cavity atom interferometer}
	
	%\title{Two-photon transitions driven by phase-coherent light resonant with the modes of a cavity atom interferometer}
	%Standing-wave effects in cavity atom interferometers with multi-chromatic light\\ What should we call our paper? --- cavity complexity completely covered

	\author{Sofus L. Kristensen}
	\altaffiliation[Present address: ]{Niels Bohr Institute, University of Copenhagen, DK-2100 Copenhagen, Denmark}
	\email{sofus.kristensen@nbi.ku.dk}
	\author{Matt Jaffe}
	\altaffiliation[Present address: ]{James Franck Institute and Department of Physics, University of Chicago, Chicago, IL, USA}
	\author{Victoria Xu}
	\author{Cristian D. Panda}
	\author{Holger M\"uller}
	\email{hm@berkeley.edu}
	%\altaffiliation[]{Corresponding author. Email:  sofus.kristensen@nbi.ku.dk (S. L. K),  hm@berkeley.edu (H. M.)}
	
	\affiliation{Department of Physics, 366 Le Conte Hall MS 7300, University of California, Berkeley, California 94720, USA}
	
	\date{\today}
	
	\begin{abstract}
		Atom interferometers in optical cavities benefit from strong laser intensities and high-quality wavefronts. The laser frequency pairs that are needed for driving Raman transitions (often generated by phase modulating a monochromatic beam) form multiple standing waves in the cavity, resulting in a periodic spatial variation of the properties of the atom-light interaction along the cavity axis. Here, we model this spatial dependence and calculate two-photon Rabi frequencies and ac Stark shifts. We compare the model to measurements performed with varying cavity and pulse parameters such as cavity offset from the carrier frequency and the longitudinal position of the atom cloud. We show how setting cavity parameters to optimal values can increase the Raman transition efficiency at all positions in the cavity and nearly double the contrast in a Mach-Zehnder cavity atom interferometer in comparison to the unoptimized case.
		%\hmu{Abstracts may have no sales pitch at all, or a little bit of it, but the emphasis should be on a disinterested description of the content of the paper. Let's rewrite the abstract at the end.}
	\end{abstract}

	\maketitle
\section{Introduction}

Cavity atom interferometers \cite{Hamilton2015CavityAI} manipulate matter waves using light coupled into an optical cavity. In recent years, cavity atom interferometers have been used for testing fundamental physics \cite{Hamilton2015, Jaffe2017} and studying forces induced by blackbody radiation \cite{Haslinger2018}. Using an optical resonator to mediate atom-light interactions \cite{Riou2017, Canuel2017, Dovale-Alvarez2017} provides clean uniform wavefronts that maintain the coherence of spatially-separated atomic wave packets, enabling interferometry times as long as 20 seconds \cite{Xu2019}. The frequencies required to drive two-photon Raman transitions for beamsplitter operations can be generated by phase modulating a single diode laser, cancelling laser noise. However, interference between the frequency components of the phase-modulated light inside the cavity leads to spatial variation in the Raman pulse efficiency, which can result in large residual ac Stark phase shifts between the beamsplitter pulses. While this variation can be suppressed by judicious choice of the single-photon detuning \cite{Wu2019} or by filtering out frequency components \cite{Dotsenko2002, Dotsenko2004}, maximizing the performance of a cavity interferometer requires a detailed model of these interference effects. Here, we model the atom-light interaction inside the cavity, and show how to use this model to improve the performance of a cavity atom interferometer. First, we theoretically describe the spatially dependent two-photon Rabi frequency and the differential ac Stark phase shift as a function of the parameters of the cavity. Then, we experimentally verify these predictions in our cavity atom interferometer \cite{Hamilton2015CavityAI, JaffeThesis2018}. We demonstrate how the transfer function of an optical cavity can be utilized to change the amplitude and phase of the intra-cavity electric fields, providing control over the constructive and destructive interference criteria. Finally, we use the model to optimize the cavity parameters and improve contrast in a Mach-Zehnder cavity atom interferometer, by increasing the Raman pulse efficiency and ac Stark phase cancellation of the beamsplitter operations.

\begin{figure}[t!]
\centering
\includegraphics[width=1\linewidth]{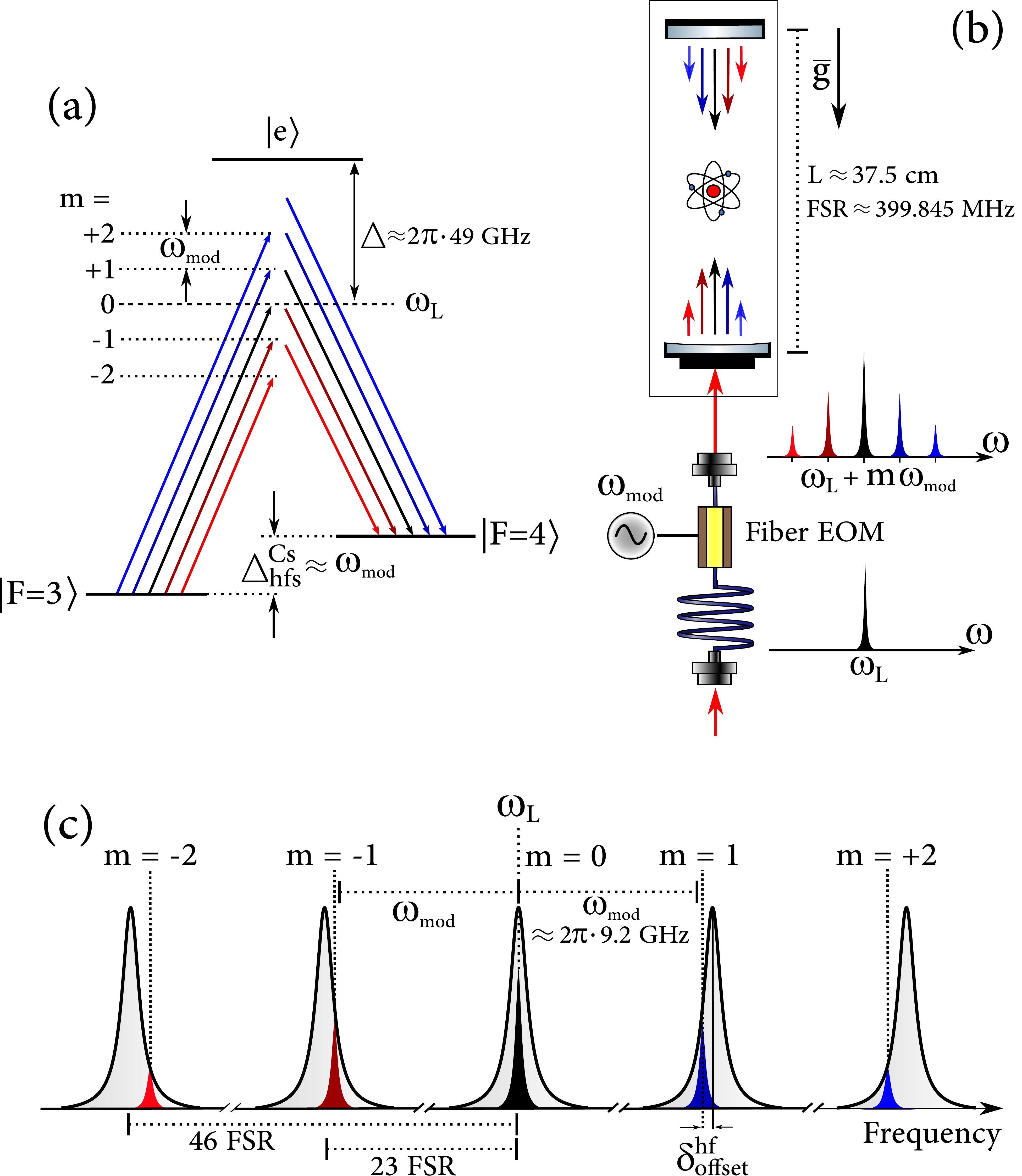} 
\caption{
(a) Cesium energy levels driven by the frequency components of a phase-modulated laser beam. 
Raman transitions are driven between the hyperfine ground states $\ket{F=3}$ and $\ket{F=4}$ via the excited state $\ket{e}$. Multiple pairs of frequency components, separated by $\omega_{\rm mod}$, simultaneously satisfy the Raman resonance. The carrier frequency, $\omega_L = \omega_{m=0}$, is red-detuned by $\Delta\approx 2\pi \cdot 49$ GHz from the $\ket{F=3}$ to $\ket{e}$ transition. (b) Spectrum of incident phase modulated light with center frequency $\omega_{\text{L}}$. The output of the fiber EOM is coupled into a vertically oriented optical cavity with length $L = 37.5 $ cm and FSR $= 399.845$ MHz. The gravitational acceleration is denoted by ${\Bar{\rm g}}$. (c) Multi-chromatic light, depicted with Lorentzians colored according to their detuning from the carrier frequency (red or blue detuned), is resonant with multiple longitudinal modes of the optical cavity, depicted with grey shaded Lorentzians. Since the linewidth of the cavity, $\gamma=2\pi\cdot3.03$~MHz, is much smaller than $\omega_\text{mod}\approx \Delta_{\text{hfs}}^{\text{Cs}}= 2\pi \cdot 9.2$ GHz necessary to drive Raman transitions in cesium, the first pair of sidebands are resonant with the cavity 23 FSR away from the carrier. 
The detuning between the first order sidebands when $\omega_\text{mod}=\Delta_{\text{hfs}}^{\text{Cs}}$ and the cavity resonance 23 FSR away is defined as $\delta_\text{offset}^{\text{hf}}$.}
\label{fig:eom_on_cavity}
\end{figure}
	
\section{Raman transitions with phase modulated light}
In our cavity atom interferometer \cite{Hamilton2015CavityAI}, beamsplitter and mirror operations are applied using a two-photon Raman transition between two hyperfine ground states of cesium, $\ket{F = 3}$ and $\ket{F = 4}$ in the $6^{2}S_{1/2}$ manifold, with frequency difference $\Delta_{\text{hfs}}^\text{Cs}=2\pi \cdot 9\,192\,631\,770$ Hz (Fig. \ref{fig:eom_on_cavity}a). The excited state $\ket{e}$ is the $6^{2}P_{3/2}$ manifold, connected to the ground state by the $852$ nm D2 line. To drive transitions between the hyperfine ground states of cesium, a fiber EOM is driven with a modulation frequency $\omega_{\text{mod}}$ near the cesium hyperfine difference frequency, $\omega_{\text{mod}} \approx \Delta_{\text{hfs}}^\text{Cs}$ (Fig. \ref{fig:eom_on_cavity}b), resulting in an output field of
\begin{align}
E_\text{out}(t) = &E_{\text{L}}e^{i\omega_{L}t + i\beta \sin\left(  \omega_{\text{mod}} t  \right)} \nonumber \\
= &E_{\text{L}}\sum_{m=-\infty}^{\infty} J_{m}(\beta) e^{i \omega_m t},
\label{eq:phase_modulation}
\end{align}
where $\omega_{m} = \omega_{\text{L}} + m \omega_{\text{mod}},$ $E_L$ and $\omega_L$ are respectively the amplitude and frequency of the incident laser beam, $J_{m}$ are the Bessel functions of the first kind, and $\beta$ is the modulation index. 
We define $\Delta$ as the single photon red-detuning (which is $2\pi \cdot 49$ GHz in our case), so that  
\begin{equation} \label{eq:omega_m}
\omega_{\text{L}}        = \omega_{3e} - \Delta ,
\end{equation}
where $\omega_{3e}$ is the resonance frequency of the $\ket{F=3}$ to $\ket{e}$ transition. On two-photon resonance,
\begin{equation} \label{eq:omega_mod}
\omega_{\text{mod}} = \Delta_{\text{hfs}}^{\text{Cs}} - \delta_{\text{ac}} - 2\delta_{\text{Dopp}}(t),
\end{equation}
where $\delta_{\text{ac}}$ is a two-photon detuning induced by the differential ac Stark shift of the two ground states, and $\delta_{\text{Dopp}}(t)$ is the Doppler shift of the atom cloud. The latter is time-dependent, as the cesium atoms are in free fall along the vertical cavity axis. To stay on two-photon resonance, we ramp the EOM modulation frequency at  $\frac{d}{dt}\omega_\text{mod}$ %over time. Cesium atoms falling due to Earth's gravitational acceleration $g$ see a Raman transition on the D2 line shift at $2\frac{d}{dt} \left( \delta_{\text{Dopp}}(t) \right) = 2 k g \approx 
$=2\pi \cdot 23.0$ $ \frac{\text{kHz}}{\text{ms}}$ during the interferometer. %, where $k=2\pi/\lambda$ is the wave number of the laser, and \lambda=852$~nm. 
For atoms freely falling for $120$ ms, $\omega_\text{mod}$ must change by $2\pi \cdot 2.76$ MHz to compensate for Doppler shift due to gravity.
	
\section{Multi-chromatic light in an optical cavity}
The phase modulated light is coupled into a cavity, as seen in Fig. \ref{fig:eom_on_cavity}b. The length $L$ of the cavity in our experiment has been adjusted such that 23 free spectral ranges (FSR) is almost equal to $\Delta^{\rm Cs}_{\rm hfs}$. Therefore, EOM sidebands separated by $\Delta^{\rm Cs}_{\rm hfs}$ can be simultaneously resonant with the cavity. We measure $L = 37.4886(2)$ cm, inferred from $\text{FSR} = 399.845(2)$ MHz.  %This is much smaller than the EOM modulation frequency, $\omega_{\text{mod}}\approx 9.2 \text{GHz}$, required to drive Raman transitions. For the Raman beams to be resonant in the cavity, $\omega_{\text{mod}}$ must be nearly an integer multiple of the cavity FSR. If this condition is satisfied, the optical frequencies separated by $\omega_{\text{mod}}$ can be simultaneously resonant with different longitudinal modes of the cavity. The cesium hyperfine splitting, $\Delta_\text{hfs}^\text{Cs}$, is nearly $23\cdot\text{FSR}$. We define the difference between these quantities as the hyperfine offset,
Because the cavity length is not aligned to a perfect integer multiple of the hyperfine frequency difference, we define the hyperfine offset

\begin{equation} \label{eq:delta_hf_offset}
\delta_{\text{offset}}^{\text{hf}} = 2 \pi \cdot 23 \cdot \text{FSR} - \Delta_{\text{hfs}}^{\text{Cs}} \approx 2\pi \cdot 3.80 \text{ MHz}.
\end{equation}
If the EOM is driven at $\Delta_{\text{hfs}}^{\text{Cs}}$ and the carrier is resonant with longitudinal mode $q$ of the cavity ($\omega_{\text{L}} = \omega_{q}$), $\omega_1$ (the first-order blue EOM sideband) is red-detuned from the cavity's longitudinal mode $q+23$ by $\delta_{\text{offset}}^{\text{hf}}$ and $\omega_{-1}$ is blue-detuned from the longitudinal mode $q-23$ by $-\delta_{\text{offset}}^{\text{hf}}$. The frequency detuning of the sidebands within the different longitudinal modes of the cavity is depicted in Fig. \ref{fig:eom_on_cavity}c. 

The cavity linewidth at $\lambda = 852$ nm is $\gamma = 2\pi \cdot 3.03(5)$ MHz. Since $\gamma, \delta_\text{offset}^\text{hf}$, and $\delta_\text{Dopp}$ are all of comparable magnitude, they must be taken into account to model the cavity atom interferometer. In addition, the laser at $\omega_L$ can be offset from cavity resonance by an amount $\delta_{\text{cav}}$,  %Where the sidebands fall on the cavity lineshape depends on the resonator's free spectral range, $\delta_{\text{cav}}$, and $\omega_\text{mod}$. 
giving sideband $m$ a frequency offset from cavity resonance of
%\begin{equation}
%\delta^{m}_{\text{cav}} = \delta_{\text{cav}} + m %\omega_{\text{mod}} - m\cdot 2\pi \cdot 23\cdot \text{FSR}.
%\end{equation}
%In our case, Eq. \ref{eq:delta_hf_offset} simplifies this expression to 
\begin{equation}
\delta^{m}_{\text{cav}} = \delta_{\text{cav}} + m \left( \delta_{\text{offset}}^{\text{hf}} + \omega_{\text{mod}} - \Delta_{\text{hfs}}^{\text{Cs}} \right).
\end{equation}
%which we will later use to calculate the phase and amplitude of the frequency components in the cavity.

The magnitude and phase of the circulating electric field within the cavity can be written in terms of these frequency components and their relative detunings. In the cavity, the upwards and downwards propagating beams form a standing wave when superimposed. This allows us to write the magnitude of the electric field in $E(z,t)$ the cavity as
\begin{align} \label{eq:multichromatic_field}
\begin{split}
E(z,t) = \sum_{j} &E_{j} \cos\left( \omega_{j} t - \varphi_{j} - k_{j}z \right) \\
&+ \sum_{l} E_{l} \cos\left( \omega_{l} t - \varphi_{l} + k_{l}z \right)
\end{split} \nonumber\\
= -2\sum_{m} &E_{m}\sin(\omega_{m}t - \varphi_{m})\sin(k_{m}z).
\end{align}
The indices $j, l$ respectively enumerate the downward and upward propagating beam. We have assumed highly reflective mirrors, so that the electric field amplitudes and frequencies of the upward and downward propagating beams can be considered equal. We denote  $k_m\equiv \omega_m/c$ (where $c$ is the speed of light), and %, $E_{j} = E_{i}$ and $\omega_{j} = \omega_{i}$. This constraint is imposed by the resonator, and is valid for mirror reflectivities close to $1$. 
%Wavenumbers $k_{m}$ are given by $k_{m} = \omega_{m} /c$ and 
$\varphi_m$ as the phase of the $m$th frequency component, %\footnote{Phase modulated sidebands are sometimes considered to have intrinsic phases. For example, the carrier and +1 sideband would have phase equal to 0, while the -1 order has $\pi$ phase (reflecting that the beat note between the carrier and the 1st-order sideband is $\pi$ out of phase with the beat note at the same frequency of the carrier and the -1st-order sideband). We carry that dependence in the sign of the Bessel function $J_{n}(\beta)$, so it does not enter into $\varphi_{i}$.} in the resonator with respect to the $m=0$ (carrier) frequency component. The phases $\varphi_{m}$ depend on the detuning from cavity resonance, as the resonator acts dispersively, imparting a frequency-dependent phase shift. That dependence is 
\cite{siegman1986lasers}
\begin{align} \label{eq:cavity_phase_txr_fxn}
\varphi_{m} &= \arg\left(\frac{\tilde{E}_{m\text{, circ}}}{\tilde{E}_{m\text{, inc}}}\right) \nonumber \\
&= \arctan\left( \frac{-r_{1} r_{2} \sin\left( \frac{\delta^{m}_{\text{cav}}}{\text{FSR}} \right) }
{1 - r_{1} r_{2} \cos\left(\frac{\delta^{m}_{\text{cav}}}{\text{FSR}} \right) } \right),
\end{align}
where $\tilde{E}_{m\text{, circ(inc)}}$ is the complex-valued circulating field of the resonator for sideband $m$, and $r_{i}$ is the reflectivity of cavity mirror $i$. 
The amplitude of each frequency component within the cavity depends on the cavity offset $\delta^{m}_{\text{cav}}$ and scales with a factor
	% Problem: We defined the k_hfs factor is only important because we have a stading wave. This is what separates our experiment from Dotsenko. We can do better than them, because they have to make an interferometer to get rid of one of the frequency components. Atom interferometry schemes such as described in Dotsenko can't use all their power.
\begin{equation} \label{eq:cavity_lorentzian}
s(\delta^{m}_{\text{cav}}) = \sqrt{\frac{\left( \gamma/2 \right)^{2}}{ {\delta^{m}_{\text{cav}}}^{2} + \left( \gamma/2  \right)^{2}} }.
\end{equation}
%normalized to unity on resonance. 
As a result, the single-photon Rabi frequency $\Omega_{m}$ for the $m$th EOM sideband is
\begin{equation}
\Omega_{m} = J_{m}(\beta)s(\delta_{\text{cav}}^{m})\Omega_{\text{L}} ,
\end{equation}
where $\Omega_{\text{L}}$ is the single-photon Rabi frequency for the full electric field strength without modulation. 
	
\section{Raman transitions in the cavity}
As shown in Appendix \ref{sec:shrodingerequation}, solving the Schr\"odinger equation that describes the interaction between the atoms and the multi-chromatic light in the cavity reveals a pattern that admits a simple expression of the Rabi frequency and ac Stark shift. The two-photon Rabi frequency $\Omega_{R}$ is given by 
\begin{equation} \label{eq:eom_rabi_nice}
4\Omega_{R}^{2} = \sum_{m, n = -N+1}^{N} R_{mn}
\end{equation}
with
\begin{equation} \label{eq:Rmn}
\begin{split}
R_{mn} &= \frac{\Omega_{m}\Omega_{m-1}\Omega_{n}\Omega_{n-1}}{(\Delta - m \Delta_{\text{hfs}}^{\text{Cs}})(\Delta - n\Delta_{\text{hfs}}^{\text{Cs}})} \\
&\times \cos\left( 2 (n-m) k_{\text{hfs}}z +\varphi_{m}-\varphi_{m-1}- \varphi_{n} + \varphi_{n-1}\right) ,
\end{split}
\end{equation}
where $\Omega_m$ is the single-photon Rabi frequency between the ground states and the excited state for frequency component $m$, and $k_{\text{hfs}} = \Delta_{\text{hfs}}^{\text{Cs}}/c$. $N$ is the highest order sidebands we include. At typical modulation depths $\beta \sim 1.2$, the electric field strength of the $m > 2$ sidebands before the cavity are small and can be neglected. 

Each term $R_{mn}$ represents the interference between a $(m,m-1)$ and a $(n,n-1)$ pair of sidebands driving Raman transitions. For $m=n$ the expression reduces to the usual two-photon Rabi frequency associated with each pair driving a Raman transition. For $m \neq n$, the Raman pairs interfere with each other constructively or destructively depending on the phases $\varphi_{i}$ of the beams, and spatial location $z$. 

%The single photon Rabi frequencies, $\Omega_m$, also depend on cavity parameters. 
Each frequency component contributes its own ac Stark shift to the hyperfine levels, which is calculated in Appendix \ref{sec:shrodingerequation} as the difference of the diagonal elements in the interaction matrix for the effective two-level system. As the single photon detuning ($\Delta \sim \text{GHz}$) is much larger than the two-photon detuning ($\delta \sim \text{kHz}$), the relevant quantity for the two-photon resonance is the differential ac Stark shift of the two ground states that shifts the two-photon resonance condition. The absolute shift of the ground states can be neglected. The total differential ac Stark shift 
\begin{equation} \label{eq:eom_stark_nice}
\delta_{\text{ac}} = \sum_{m=-N}^{N} S_{m},
\end{equation}
is found to be the sum of the ac Stark shifts of the individual frequency components $m$,
\begin{equation} \label{eq:Sm}
S_{m} = \frac{\Omega_{m}^{2}}{4} \left(  \frac{1}{\Delta - (m+1) \Delta_{\text{hfs}}^{\text{Cs}}}  -  \frac{1}{\Delta - m \Delta_{\text{hfs}}^{\text{Cs}}}  \right).
\end{equation}

\begin{figure}
\centering
\includegraphics[width=1\linewidth]{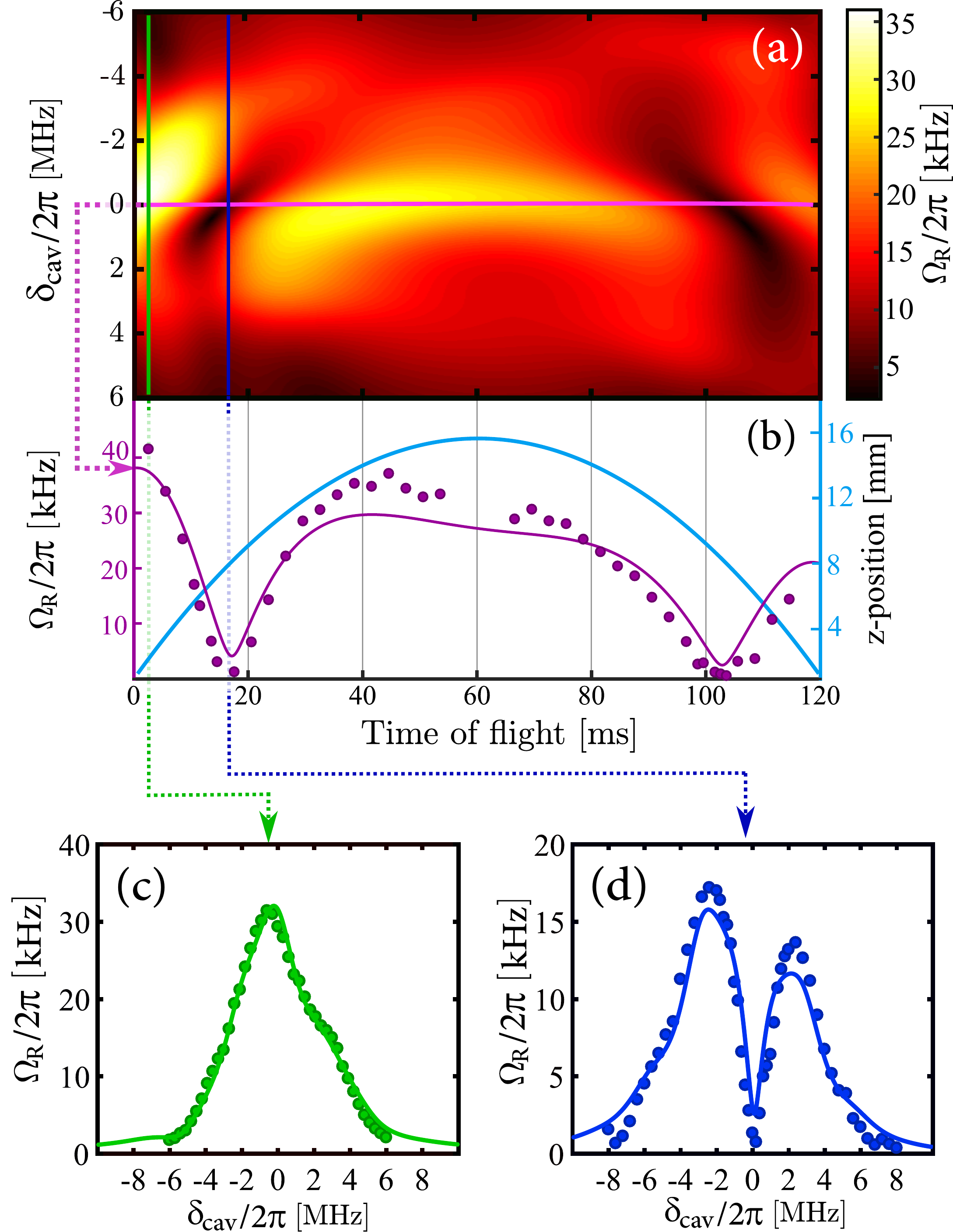}
\caption{(a) Simulation of the Rabi frequency, $\Omega_{R}$, versus $\delta_{\text{cav}}$ and time of flight. The colored lines correspond to the data presented in Fig. (b), (c) and (d). (b) On the left axis, Rabi frequency versus time of flight. The spatial beat note of the Raman beam pairs can be seen, as well as a general downward trend as $\omega_{\text{mod}}$ is linearly ramped to account for the Doppler shift, moving the sidebands out of the cavity lineshape resulting in lower light intensity in the cavity. On the right axis, the atom cloud position is plotted as a function of time of flight. The cloud passes the same Rabi dead zone twice. (c), (d) Measured Rabi frequency as a function of cavity offset compared to simulation, near a (c) maximum ($t = 2.0$ ms) and (d) minimum ($t = 16.5$ ms) of the $\delta_{\text{cav}}$ = 0 MHz spatial beat note, respectively.}
\label{fig:Rabi_frequencies2}
\end{figure}

Together, Eqs. \ref{eq:eom_rabi_nice} and \ref{eq:eom_stark_nice} with the definitions of $\omega_{\text{mod}}$, $\delta_\text{cav}^m, \varphi_m$ and $\Omega_m$, predict the Rabi frequencies and light shifts for cesium atoms in our optical cavity. This is the model we set out to find, and in the next section, we will compare our theoretical description of intra-cavity Raman transitions with in-situ measurements performed in our cavity atom interferometer (Figs. \ref{fig:Rabi_frequencies2} and \ref{fig:acstarkmethod}).

%\hmu{We should add a sentence or two highlighting that this is the model we were looking for. We should point to the graphs in Figure 2 that show these predictions.}
	
% Near the apex, when the atoms are close to zero velocity, we cannot distinguish between upward and downward moving beams, and so we are unable to drive Raman transition with a specific direction of the photon momentum kick. \slk{fix this. Maybe remove apex}

\section{Measuring of the two-photon Rabi frequency and the ac Stark shift} 
To test the predictions of our model, we have measured the Rabi frequency of the Raman transition and the differential ac Stark shift of the two ground states in our cavity atom interferometer. Our setup has been described in \cite{Hamilton2015CavityAI}. We laser cool cesium atoms and launch them upwards into free-fall along the cavity mode. While the atoms are freely falling, we apply laser pulses and measure the two-photon Rabi frequency $\Omega_R$ and ac Stark shift $\delta_\text{ac}$ as a function of parameters such as the cavity-laser detuning and free-fall time after the launch.
	
\subsection{The two-photon Rabi frequency}
Figure \ref{fig:Rabi_frequencies2}a shows a simulation of the Rabi frequency in our cavity as a function of the time of flight of the atom cloud and the cavity offset. This is calculated from Eqs. (\ref{eq:eom_rabi_nice}, \ref{eq:eom_stark_nice}), by inserting the atom position as function of free-fall time and the appropriate time-dependent $\delta_\text{Dopp}(t)$. The colored lines indicate times at which we compare the measured Rabi frequency and cavity offset to theory.

To measure the Rabi frequency, we drive Rabi oscillations using a square pulse of varying duration generated by an acousto-optic modulator (AOM) and measure the probability that the atom undergoes a transition. For comparison with theory, we fit Rabi oscillations of the atomic cloud to a model which accounts for the cloud’s thermal expansion, and extract the peak Rabi flopping of atoms at the center of the cavity mode $\Omega_R$ from our fit. A complete description of how $\Omega_R$ is extracted from a fit to the Rabi oscillations of the ensemble can be found in Appendix \ref{sec:fit_appendix}. This more complex fit model is required because the cavity beam waist is roughly the size of the atom cloud.

%From the fit we extract $\Omega_R$, defined as the two-photon Rabi frequency of an atom that experiences the field intensity in the center of the cavity mode. The complete description of the fit model can be found in Appendix \ref{sec:fit_appendix}. 
% \hmu{does the model also take into account the velocity distibution?
In Fig. \ref{fig:Rabi_frequencies2}b we present the measured two-photon Rabi frequency $\Omega_{R}$ as a function of free-fall time at $\delta_{\text{cav}} = 0$. Different free-fall times correspond to different locations along the the cavity axis, and trajectory of the atoms is indicated on the second y-axis of Fig. \ref{fig:Rabi_frequencies2}b. The general decreasing trend of the Rabi frequency over time is caused by $\omega_{\text{mod}}$ being linearly ramped to stay on two-photon resonance with the falling atoms. This causes the cavity detuning of the sidebands to increase, reducing the amplitude of the intracavity field over time (see Eq. \ref{eq:cavity_lorentzian}). %From Eq. \ref{eq:Rmn}, we see the $\cos\left( 2 k_{\text{rf}} z  + \ldots \right)$ factor give rise to interference via the $R_{01}$ and $R_{10}$ terms \hmu{There is no such factor in Eq. (9). Can you be more specific?}. 
For the modulation depth used ($\beta = 1.08$), the carrier and first order sidebands dominate, as higher order sidebands out of the EOM are weak and almost entirely suppressed by the cavity. As a result, the dominating spatial periodicity of the interference is $2\pi/k_{\text{hfs}} \simeq 1.63$\,cm, which is comparable to the distance the atoms move over 120 ms in free fall. The low Rabi frequencies at $\approx 17$ ms and $\approx 103$ ms show atoms passing the same region of destructive interference between the Raman pairs twice, since the atoms reach the apex of their trajectory after 60 ms.

Figs. \ref{fig:Rabi_frequencies2}c and \ref{fig:Rabi_frequencies2}d show the Rabi frequency as a function of cavity offset for atoms at two locations: an amplitude maximum and minimum, respectively, of the Rabi frequency spatial beat note with $\delta_{\text{cav}} = 0$ MHz. Near the minimum, the Rabi frequency can be increased by detuning the laser from cavity resonance in either direction. This alleviates the destructive interference by changing the relative phase of the interfering beams (Eq. \ref{eq:cavity_phase_txr_fxn}), while also emphasizing one sideband over the other within the cavity lineshape (Eq. \ref{eq:cavity_lorentzian}). 

%The dependence of $\Omega_{R}$ on $\delta_{\text{cav}}$ at a maximum of the spatial beat note (Fig. \ref{fig:Rabi_frequencies2}c) is significantly different than at a minimum of the beat note (Fig. \ref{fig:Rabi_frequencies2}d), i.e. the Rabi dead zone.

%Near the minimum, the Rabi frequency can be increased by detuning the laser from cavity resonance in either direction.

%the cavity transfer function changes the relative phase of the interfering beams (Eq. \ref{eq:cavity_phase_txr_fxn}) to alleviate the destructive interference, while also emphasizing one sideband over the other within the cavity lineshape (Eq. \ref{eq:cavity_lorentzian}).
	
\begin{figure} 
\centering
\includegraphics[width=1.0\linewidth]{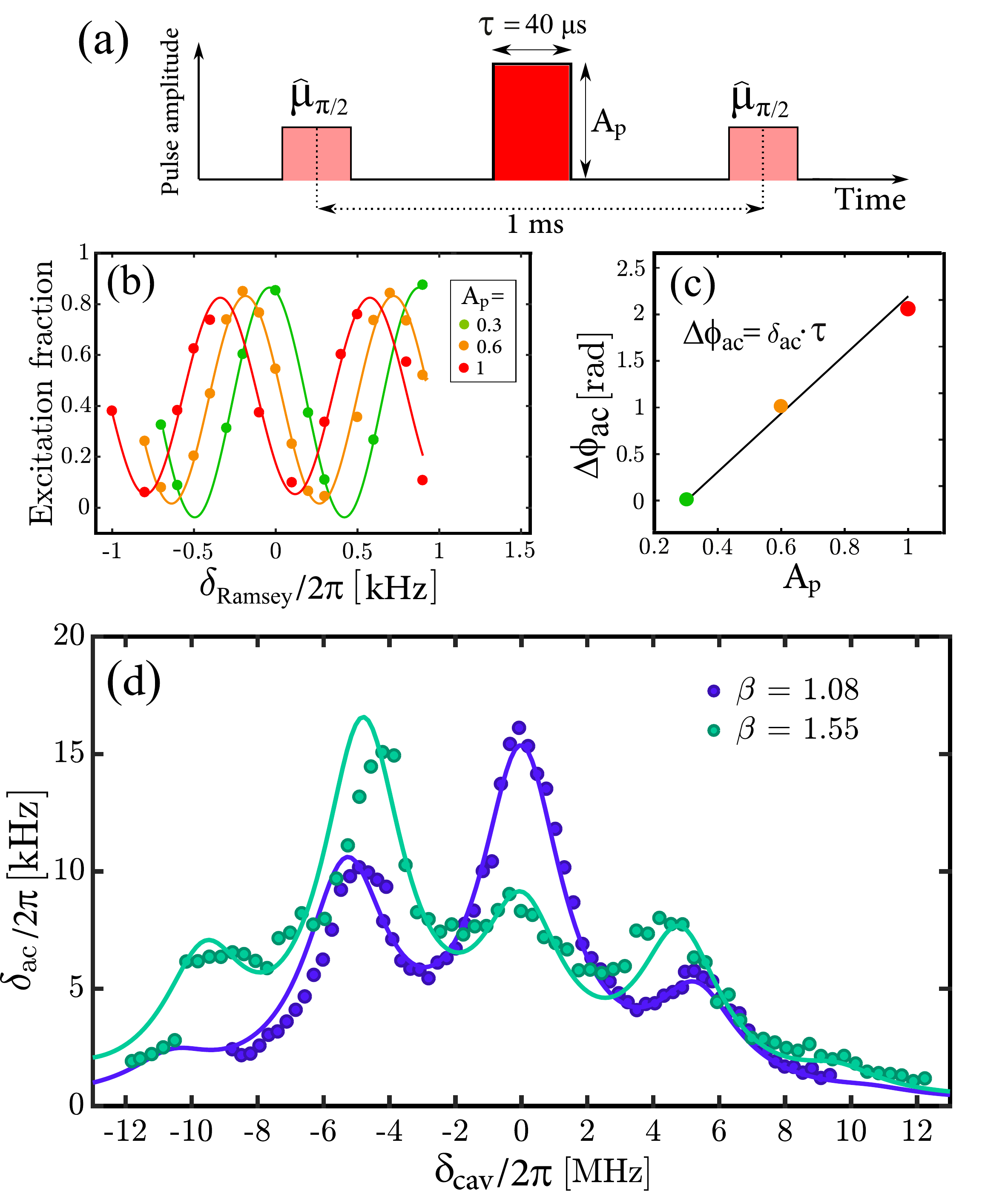}
\caption{(a) Pulse sequence of the in-situ ac Stark shift measurement using a microwave Ramsey interferometer. %\hmu{Label the y axis in the figure. I took out the next sentence as it is a repetiton of what is said in the text.} %Two microwave $\pi/2$-pulses, denoted by $\hat{\text{\textmu}}_{\pi/2}$, open and close the interferometer, while a $40$-\textmu s long ac Stark pulse phase shifts the interferometer arms. 
(b) Ramsey fringes measured for different pulse amplitudes. The frequency of the microwave $\pi/2$-pulses is scanned by an amount $\delta_\text{Ramsey}$ to obtain interference fringes which accrue phase at a rate of $2\pi/T$, as observed in the population oscillations of atoms between states $\ket{F=3}$ and $\ket{F=4}$. %Increasing the pulse amplitude shifts the phase of the fringes. 
(c) Phase shift, in radians, as a function of pulse amplitude, $A_\text{p}$, in arbitrary units. The ac Stark phase shift, measured as the phase offset of fringes in (b), increases linearly with the pulse amplitude, at a rate determined by the duration of the ac Stark pulse $\tau$, and the ac Stark energy level shift $\Delta \phi_\text{ac} = \delta_\text{ac}\cdot \tau$. 
(d) ac Stark shift measured as a function of cavity offset $\delta_{\text{cav}}$ for two different modulation depths $\beta$. Different modulation frequencies were used as well, as seen from the sideband peak locations ($\omega_{\text{mod}} = \Delta_{\text{hfs}}^{\text{Cs}} - 2\pi \cdot 1.500\,[1.000]$ MHz for $\beta = 1.08\,[1.55]$).}
		\label{fig:acstarkmethod}
\end{figure}
%The ac Stark shift does not depend on axial position $z$, as the size of the atom cloud is much bigger than the wavelength of the electric field.}
%The phases of interference fringes from (b) are fitted to a straight line as a function of pulse amplitude $A_\text{p}$. The slope of this line is decided by the length of the ac Stark pulse $\tau$, and the ac Stark shift $\delta_\text{ac}$. From this we can determine $\delta_\text{ac}$.

\subsection{The ac Stark shift}

In this section, we describe our measurement of the ac Stark shift in the cavity for different cavity and modulation parameters. Contrary to the Rabi frequency, the ac Stark shift $\delta_\text{ac}$ has no spatial dependence because it has no interference terms, as seen in Eq. \ref{eq:Sm}. However it does depend on the cavity offset, $\delta_\text{cav}$, and modulation depth, $\beta$. %, as changing the cavity offset accentuates different frequency components of the free-space spectrum, and $\beta$ decides their initial strengths in free-space. Consequently, it is relevant to study the ac Stark dependence on $\delta_\text{cav}$ for different modulation depths $\beta$.

We measure the ac Stark shift of a given laser pulse by using the microwave Ramsey procedure \cite{Peters1998Thesis} depicted in Fig. \ref{fig:acstarkmethod}a. A microwave $\pi/2$-pulse with frequency $\Delta^\text{Cs}_\text{hfs}-\delta_{\rm Ramsey}$ puts the atoms into an equal superposition of $\ket{F=3}$ and $\ket{F=4}$. The system evolves at a rate of $\Delta_{\text{hfs}}^\text{Cs}$ for a time $T_{\rm Ramsey}=1$ ms after which we apply another microwave $\pi/2$-pulse to close the Ramsey interferometer. A Raman pulse of duration $\tau=40$ \textmu s is applied during the Ramsey time $T_{\rm Ramsey}$ to impart an ac Stark phase shift of $\Delta \phi_\text{ac} = \delta_{\text{ac}}\cdot \tau$ on the atoms. We detune $\omega_{\text{mod}}$ from Raman resonance by $2\pi \cdot 1$ MHz (for $\beta = 1.08$) and $2\pi \cdot 1.5$ MHz (for $\beta = 1.55$) such that laser pulse does not drive transitions, and we measure interference fringes by varying $\delta_\text{Ramsey}$. The different detunings are chosen to increase the visibility of ac Stark shift induced by the higher order sidebands for a higher modulation depth. The phase of the interference signal is obtained by fitting a sine to the excitation fraction as a function of $\delta_\text{Ramsey}$ (Fig. \ref{fig:acstarkmethod}b). The variation of the ac Stark-induced phase shift with the amplitude of the ac Stark Raman pulse $A_\text{p}$ (in arbitrary units) is shown in Fig. \ref{fig:acstarkmethod}c. 

We extract the ac Stark energy level shift $\delta_\text{ac}$ in Fig. \ref{fig:acstarkmethod}d by measuring how the light-induced phase shift $\Delta\phi_\text{ac}$ varies with pulse amplitude, for a fixed pulse duration $\tau$; that is, the measurements of $\delta_{\rm ac}$ in Fig. \ref{fig:acstarkmethod}d correspond to the slope measured in Fig. \ref{fig:acstarkmethod}c, divided by the ac Stark pulse width $\tau$.

%At least two data points are needed to quantify the ac Stark shift. The third data point is included as a linearity check. The offset of the data points have been adjusted to zero the first data point, as the relevant quantity is the slope of the $A_\text{p}$ vs. $\Delta \phi_{\rm ac}$ linear fit. By substituting in our usual amplitude $A_p=1$ for interferometer pulses, and the duration of the ac Stark pulse $\tau$, we find $\delta_{\text{ac}}$ by dividing the accumulated ac Stark phase, $\Delta \phi_\text{ac}$, with the ac Stark pulse time, $\tau$.

Fig. \ref{fig:acstarkmethod}d shows strong agreement between the measured and predicted ac Stark shifts for two modulation depths $\beta$. The only free parameter in the theory curve is the amplitude scaling (as the overall amplitude is unimportant). At high modulation depth (shown in green), second-order sidebands become visible as additional peaks in the ac Stark shift spectrum. In addition, the positive (blue) sidebands are observed to cause stronger ac Stark shifts. This is expected at negative cavity detuning, as the blue sidebands have a smaller detuning to the single-photon transition (see Fig. \ref{fig:eom_on_cavity}).
%The small deviation from the model may be caused by the slow heating of the vacuum chamber which houses cavity which slightly changes the FSR of the cavity during long operation. \slk{make previous sentence a little less obvious in your face}

%\hmu{In the end, double check that figures are located on the page where they are first referenced in the text or on a later page, but not on an earlier page.}
	
\section{Increasing contrast of a Mach-Zehnder atom interferometer}
We now use the model to improve the performance of a Mach-Zehnder atom interferometer, by exploiting how the cavity parameters change the Rabi frequency and ac Stark shift. In a Mach-Zehnder atom interferometer, the ac Stark shifts applied during the beamsplitter pulses should ideally cancel \cite{Weiss1994, Peters1998Thesis}. However, a sensitive Mach-Zehnder atom interferometer requires a long pulse separation time $T$, which requires the %a sensitive Mach-Zehnder atom interferometer requires a long pulse separation time $T$, which requires the 
atom-light interactions to happen at different locations along the cavity axis, between which the Rabi frequency $\Omega_{R}$ can vary greatly. Since the duration $\tau$ of a $\pi/2$ pulse is $\tau = \frac{\pi}{2}\frac{1}{\Omega_{R}}$, the ac Stark phase shift resulting from such a beamsplitter pulse is $\varphi^{\text{ac}} = \frac{\pi}{2} \frac{\delta_{\text{ac}}}{\Omega_{R}}$. Making this phase shift equal for both beamsplitter pulses thus requires detailed understanding of the Rabi frequency and ac Stark shift.

To optimize the interferometer performance, we engineer the interferometer to 1) cancel the ac Stark shift phase between pulses and 2) minimize the ac Stark phase added to the interferometer by each pulse. %This quantity should be equal for both beamsplitter pulses, which both reduces the systematic ac Stark phase shift, and increases the contrast by suppressing the spread in ac Stark shift across the cloud. Additionally, the individual $\varphi^{\text{ac}}$ of each beamsplitter pulse should be small, 
The latter helps, as cloud expansion and other effects make perfect cancellation of the residual total ac Stark phase shift $\Delta \varphi_{\rm res}^{\text{ac}} := | \varphi^{\text{ac}}_{1} - \varphi^{\text{ac}}_{3}|$ impossible for all atoms simultaneously. In particular, applying interferometer pulses at locations where the Rabi frequency is low (the Rabi dead zone) necessitates more powerful or longer pulses to realize beamsplitters; combined with the inhomogenous addressing of a thermal atom cloud, these lower Rabi frequencies can result in a large residual light shift between the first and final pulses that significantly reduces interferometer contrast. 
	
%To demonstrate this, we run a Mach-Zehnder interferometer with $T = 10$  ms, where the three pulses occur at $20, 30,$ and $40$ ms after launch (as seen in Fig \ref{fig:interferometer}a). The first beamsplitter pulse is applied close to the amplitude minimum of the Rabi beat note, while the third beamsplitter occurs at a maximum, which results in poor ac Stark phase shift matching. For this extreme configuration, the calculated ac Stark phase shift for a $\pi/2$ pulse is plotted as a function of cavity offset and time of flight in Fig. \ref{fig:interferometer}b.
%For $\delta_\text{cav}=0$, the phase shift of the first and third pulses, indicated by the dotted lines, has poor cancellation and leaves over 1 radian of residual phase shift in the interferometer.

\begin{figure}
\centering
\includegraphics[width=0.95\linewidth]{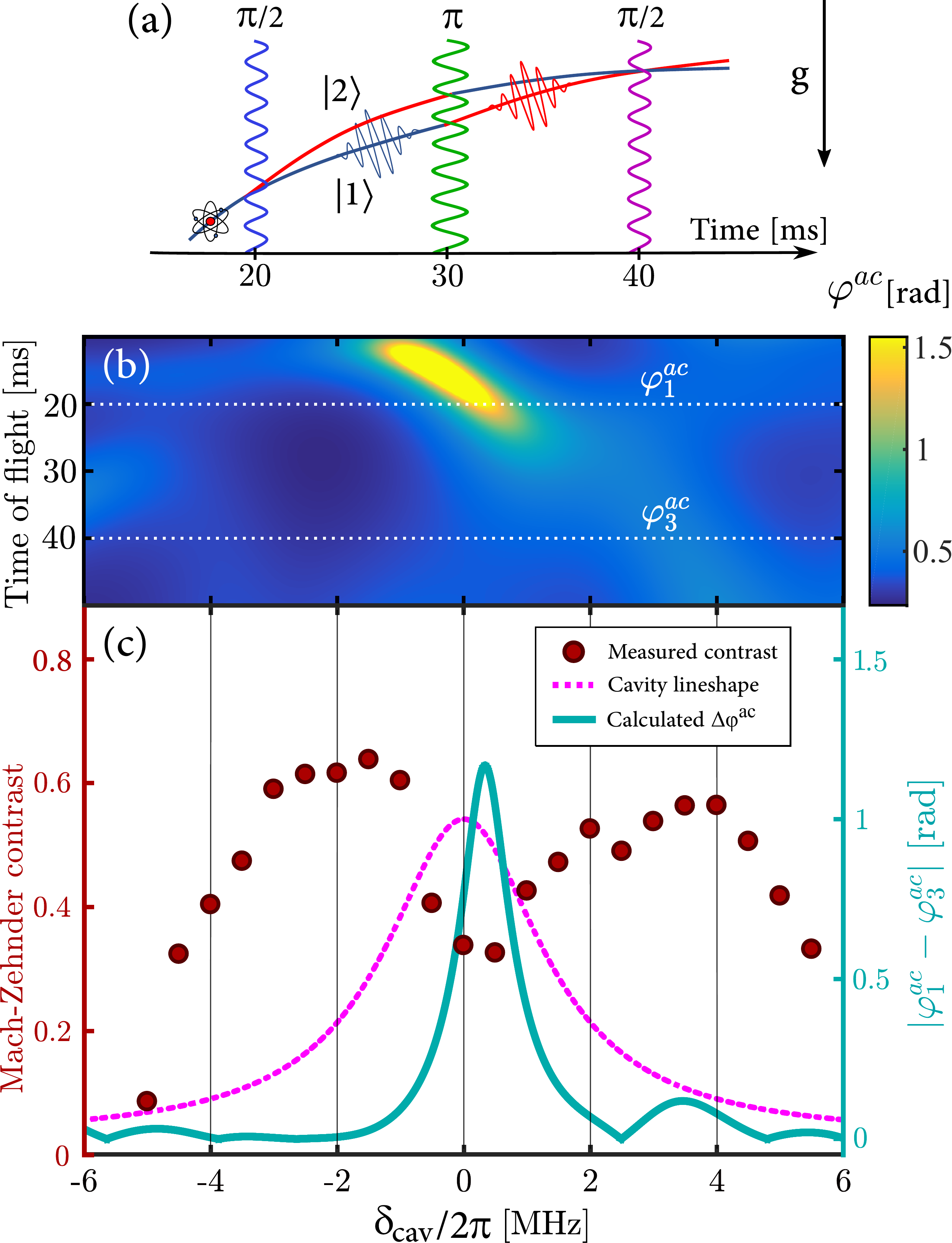} 
\caption{ (a) Schematic of the Mach-Zehnder atom interferometer with $T=10$\,ms.  Atoms fall along the cavity axis ($g$ denotes the direction of gravity), and $T$ was chosen such that the beamsplitter pulses occur at different extrema along the spatial Rabi beat note (see Fig. \ref{fig:Rabi_frequencies2}b).
(b) Theoretical ac Stark phase shift $\varphi^{\text{ac}} = \frac{\pi}{2} \frac{\delta_{\text{ac}}}{\Omega_{R}}$ of a $\pi/2$-pulse as a function of cavity offset and time of flight. The first $\pi/2$-pulse at 20 ms is near the Rabi dead zone for $\delta_{\text{cav}}=0$, causing it to induce a large phase shift as the pulse is long. However, by changing $\delta_\text{cav}$ for all pulses it is possible to reduce $\varphi^{\text{ac}}_{1}$ to match $\varphi^{\text{ac}}_{3}$ more efficiently. (c) Interferometer contrast vs. cavity offset $\delta_{\text{cav}}$. The contrast measurements are plotted with red data points, while the solid cyan line represents the calculated residual ac Stark shift phase from the first and third beamsplitter pulses $|\varphi_1^\text{ac}-\varphi_3^\text{ac}|$. In dashed purple the normalized cavity lineshape $s(\delta_{\text{cav}})^2$ is plotted for $\delta_{\text{cav}}=0$. We observe that the contrast decreases significantly at $\delta_{\text{cav}}\approx 0$, where $|\varphi_1^\text{ac}-\varphi_3^\text{ac}|$ is large.}
\label{fig:interferometer}
\end{figure}

To demonstrate this, we run a Mach-Zehnder interferometer where the first beamsplitter pulse is applied close to the amplitude minimum of the Rabi beat note, while the third beamsplitter occurs at a maximum, which results in poor ac Stark phase shift matching. This corresponds to a Mach-Zehnder with $T = 10$ ms, where the three pulses occur at $20$, $30$, and $40$ ms after launch, as shown in Fig. \ref{fig:interferometer}a. For this extreme configuration, the calculated ac Stark phase shift for a $\pi/2$ pulse is plotted as a function of cavity offset and time of flight in Fig. \ref{fig:interferometer}b.
For $\delta_\text{cav}=0$, the phase shift of the first and third pulses, indicated by the dotted lines, has poor cancellation and leaves over 1 radian of residual phase shift in the interferometer. 

Nonetheless, the interferometer performance can be recovered by offsetting the cavity detuning when applying interferometry pulses to modify the Rabi frequency and the ac Stark shift. In Fig. \ref{fig:interferometer}c, we present the measured contrast as a function of $\delta_\text{cav}$ alongside the calculated residual phase shift $\Delta \varphi_{\rm res}^{\text{ac}}$. This shows how the contrast increases when $\Delta \varphi_{\rm res}^{\text{ac}}$ is minimized. Also shown in dashed purple is the cavity lineshape, indicating the relative amplitude of the carrier frequency when varying the cavity offset. The fringe contrast was improved from $\approx 34\%$ with $\delta_{\text{cav}} = 0$ (where $\Delta \varphi_{\rm res}^{\text{ac}}$ is large) to over $60$\% by setting $\delta_{\text{cav}}$ near $-2\pi \cdot 2$ MHz to minimize $\Delta \varphi_{\rm res}^{\text{ac}}$, as predicted by the model. %This can prove vital for measurements where the atomic trajectory must be tuned to maximum sensitivity towards a specific potential gradient.
This shows how the cavity can be utilized as a tool to recover and optimize interferometry pulses at arbitrary locations along the cavity axis, despite the spatially varying Rabi frequency that results from using phase-modulated light to drive Raman transitions.
	
%With this, we show that the contrast can nearly double for cavity offsets which 1) cancel the ac stark shift phase between pulses, and 2) minimize the total ac stark phase added to the interferometer by each pulse. For cavity offsets larger than $\pm 4$ MHz, the performance of the interferometer drops as the carrier frequency is ramped too far away from cavity resonance. 
	
	%We posit that varying the cavity offset between each interferometer pulse could also result in improved contrast, as $\delta_\text{cav}$ could be tuned during the interferometer sequence to maximize the Rabi frequency for the different time of flights.

\section{Conclusion}
We have developed and experimentally confirmed a model of intra-cavity Raman pulses driven by phase modulated light. In turn, this model has allowed us to nearly double the contrast of a Mach-Zehnder cavity atom interferometer. % by mapping the parameter space that decides the Rabi frequency and ac Stark shift, and it may prove useful for future cavity atom interferometer implementations. 
The cavity provides intensity enhancement and mode-filtering of the interferometer beam, which has been instrumental in achieving 20-s hold times in a lattice interferometer \cite{Xu2019}, but it also makes the Rabi frequency of beamsplitter pulses spatially dependent, which can affect the contrast of the interferometer and its systematic effects. We have shown that these problems can be overcome by using the cavity resonance to adjust the amplitude and phase of the electric field components, to maximize Rabi frequency at necessary locations while minimizing the ac Stark shift between the pulses.%This has the advantage over free-space atom interferometers that it allows easy control of the spatial interference of the Raman transitions. 

The freedom to perform interferometry pulses at arbitrary locations along the cavity axis is critical for measurements where the atomic trajectory has been optimized to measure a specific potential gradient. This is often needed in precision measurements based on atom-source mass interactions, such as in searches for screened dark energy candidates \cite{Hamilton2015, Elder2016, Jaffe2017}, observing a gravitational analogue of the Aharanov-Bohm effect \cite{gravAB}, or measurements of the gravitational constant $G$ \cite{Fixler2007,Lamporesi2008,Rosi2014}. 

The techniques demonstrated in this work will be relevant when implementing new ideas in cavity atom interferometry, such as using higher-order transverse modes for more sophisticated optical traps, performing high-order Bragg diffraction for large momentum transfer atom optics \cite{Muller2008BraggTheory}, or incorporating spin squeezing of atoms in optical cavities \cite{Hosten2016, Cox2016} for quantum-enhanced atom interferometry.
	
\section{Acknowledgements}
We would like to thank Philipp Haslinger for contributions to the experimental apparatus and for reviewing the manuscript. Additionally, we wish to acknowledge Paul Hamilton for initially discovering and stating the problem of Raman transition interference.

This material is based on work supported by the National Science Foundation under grant no. 1708160, the National Aeronautics and Space Administration grants no. 1629914 and 1645921 and the Office of Naval Research grant no. N00014-20-1-2656
\appendix

\section{Solving the three-level system for a multi-chromatic standing wave} \label{sec:shrodingerequation}
We wish to derive the Rabi frequency of the two-photon Raman transition between two hyperfine ground states, here generalized as $\ket{1}$ and $\ket{2}$, of a three-level atomic system, as depicted in Fig. \ref{fig:eom_on_cavity}c. Our derivation will follow Refs. \cite{Dotsenko2002, Dotsenko2004}, who consider two-photon Rabi oscillations driven by one pair of Raman resonant sidebands, resulting in two interfering Raman transitions. We extend these results to describe a standing wave in an optical cavity consisting of $N$ frequency components and $N-1$ interfering frequency pairs on Raman resonance. 

We write the state vector $\hat{\psi}(t)$ as a superposition of the eigenstates of the Hamiltonian,
\begin{equation} \label{eq:prob_amp}
    \hat{\psi}(t) = \sum_{i} C_i(t) e^{-i\Phi_i (t)} \ket{i}
\end{equation}
\noindent
for $i\in \{1,2,e\}$. The $C_i$ are real probability amplitudes, and $\Phi_i (t)$ are time-dependent phases of the states.
The Hamiltonian for the full three-level system is
\begin{align}
\hat{H}(t) =
    \begin{pmatrix}
E_1^0 & 0 & V_{e1}(t) \\
0 & E_2^0 & V_{e2}(t) \\
V_{1e}(t) & V_{2e}(t) & E_e^0
\end{pmatrix}
\end{align}
where $E_i^0$ are the ground state energies of the levels, and
\begin{equation} \label{eq:dipoleint}
    V_{ij}(t) = V_{ji}(t) = -\bold{d}\cdot \bold{E}(t)
\end{equation}
is the interaction between the dipole moment $\bold{d}$ and the electric field $\bold{E}(t)$. The elements coupling states $\ket{1}$ and $\ket{2}$ are $H_{12}=H_{21}=0$ due to orthogonality (no direct transition between these states). In order to determine the other matrix elements, we write the single photon Rabi frequency for the $m$th frequency component as
\begin{equation} \label{eq:single_photon_Rabi}
\Omega_{m,i} = \frac{1}{\hbar} \bra{i}\boldsymbol{d} \cdot \boldsymbol{E}_{m} \ket{e},
\end{equation}
\noindent
where $i=1,2$. We assume that $\Omega_{m,1}=\Omega_{m,2}$. Using Eqs. (\ref{eq:dipoleint}, \ref{eq:single_photon_Rabi}) and the electric field Eq. \ref{eq:multichromatic_field}, we obtain  
\begin{equation}
    V_{1e}(t)=V_{2e}(t) = 2 \sum_{m=-N}^{N} \hbar \Omega_{m}  \sin(\omega_{m} t- \varphi_{m})\sin(k_{m}z).
\end{equation}
We will take $\Omega_{m}$ to be real, and absorb any complex phase in to the electric field phase $\varphi_{m}$. 
To find differential equations for the probability amplitudes in Eq. (\ref{eq:prob_amp}), we start from the  Schr\"odinger equation,
\begin{align}
\begin{split}
    i\hbar \frac{d}{dt}C_i(t) =&[E^0_i -\hbar \dot{\Phi}_i(t)] C_i(t) \\ &+ \sum_{j=1}^3 V_{ij} C_i(t)e^{i( \Phi_i(t) - \Phi_j(t)  )},
\end{split}
\end{align}
where we define the time-dependent state phases $\dot{\Phi}_i$ as 
\begin{equation} \label{eq:statephases}
\begin{split}
    \dot{\Phi}_1 &= E_1^0/\hbar \\
    \dot{\Phi}_2 &= E_2^0/ \hbar - \delta_\text{ac} \\
    \dot{\Phi}_e &= E_e^0 /\hbar,
\end{split}
\end{equation}
where $\delta_\text{ac}$ is a frequency shift of $\ket{2}$ that we later will solve for in order to extract the differential light shift of the two ground states. The differential phases then become,
\begin{equation}
    \begin{split}
        \Phi_1-\Phi_e = \frac{1}{\hbar} ( E_1^0 - E_e^0 )t =& -\omega_{1e}t \\
        \Phi_2-\Phi_e = \frac{1}{\hbar} ( E_2^0 - E_e^0 )t =& -\omega_{2e}t-\delta_\text{ac} t
    \end{split}
\end{equation}
where $\omega_{1e}$ and $\omega_{2e}$ are the resonance frequencies from $\ket{1}$ $\rightarrow$ $\ket{e}$, and $\ket{2}$ $\rightarrow$ $\ket{e}$, respectively, which are given by
\begin{align}
\omega_{1e}&=\omega_m+\Delta-m(\Delta_\text{hfs}^\text{Cs}+\delta_\text{ac}) \\
\omega_{2e}&=\omega_{1e}-\Delta_\text{hfs}^\text{Cs}+\delta_\text{ac}.
\end{align}
Using the definitions of phases in Eq. \ref{eq:statephases} removes the unperturbed energies from the diagonal elements of the Hamiltonian. The probability amplitudes therefore satisfy 
\begin{align}
        i\hbar \frac{d}{dt}C_1(t) &= V(t) C_e(t)e^{-i\omega_{1e}t} \label{eq.c1} \\
        i\hbar \frac{d}{dt}C_2(t) &=\delta_\text{ac} C_2 + V(t) C_e(t)e^{-i(\omega_{2e}+\delta_\text{ac})t} \label{eq.c2}\\
        i\hbar \frac{d}{dt}C_e(t) &= V(t) C_2(t)e^{i(\omega_{2e}+\delta_\text{ac})t} + V(t) C_1(t)e^{i\omega_{1e}t} \label{eq.c3}.
\end{align}

We now adiabatically eliminate the excited state to describe the system as an effective two-level system. Since $\Delta$ is much larger than the linewidth of the single photon transition, the population of the excited state will be small and varying quickly. We can therefore integrate Eq. (\ref{eq.c3}), assuming $C_1$ and $C_2$ to be constant,% in order to obtain an expression for $C_e$ in terms of $C_1$ and $C_2$,
\begin{align}
        i\hbar C_e(t) &=  C_2 \int dt V(t)  e^{i(\omega_{2e}+\delta_\text{ac})t} + C_1 \int dt V(t)  e^{i\omega_{1e}t} \label{eq:integratec3}.
\end{align}
We can use this to eliminate $C_e$ from Eqs. (\ref{eq.c1}) and (\ref{eq.c2}).% such that $C_1$ and $C_2$ have no dependence on $C_e$, and the system has been reduced to an effective two-level system.

As the final step, we apply the rotating wave approximation (RWA) by eliminating any terms oscillating faster than the evolution of the state population, i.e., terms oscillating at a sum frequency. %We apply the RWA, make substitutions and reduce the equations. 
The system we wish to solve can now be written as
\begin{equation} \label{eq:generaltwolevel}
    \begin{pmatrix}
    \dot{C_1} \\ \dot{C_2}
    \end{pmatrix} = \frac{1}{4}
    \begin{pmatrix}
    A & B \\
    C & D \\
    \end{pmatrix}
    \begin{pmatrix}
    C_1 \\ C_2
    \end{pmatrix},
\end{equation}
where
\begin{align}
A&= \sum_{m=-N}^{N}\frac{\Omega_{m}^{2}}{\Delta - m \Delta_{\text{hfs}}^{\text{Cs}}}, \label{eq:A} \\
    B&= \sum_{m=-N+1}^{N} \frac{\Omega_{m-1} \Omega_{m} e^{i((k_{m-1}+k_{m})z+\phi_{m-1}-\phi_{m})}}{(\Delta-m\Delta^\text{Cs}_\text{hfs})},\label{eq:B} \\
    C&= \sum_{m=-N+1}^{N} \frac{\Omega_{m-1} \Omega_{m} e^{-i((k_{m-1}+k_{m})z+\phi_{m-1}-\phi_{m})}}{(\Delta-m\Delta^\text{Cs}_\text{hfs})},\label{eq:C} \\
    D&= \sum_{m=-N}^{N}\frac{\Omega_{m}^{2}}{\Delta - (m+1) \Delta_{\text{hfs}}^{\text{Cs}}}-4\delta_\text{ac}. \label{eq:D}
\end{align}
If the atoms starts in $\ket{1}$ (i.e. $C_{1}(t=0)=1$, $C_{2}(t=0)=0$), the two-level system described in Eq. $\ref{eq:generaltwolevel}$ oscillates between states $\ket{1}$ and $\ket{2}$ as

	\begin{align}
	|C_{1}(t)|^2 &= 1 - \Lambda \sin^{2}\left( \frac{\Omega_{R}}{2} t \right)  \\
	|C_{2}(t)|^2 &= 	\Lambda \sin^{2}\left( \frac{\Omega_{R}}{2} t \right)
	\end{align}
	\noindent
	with
	\begin{align}
	\Lambda = \frac{4BC}{(A - D)^{2} + 4BC}
	\end{align}
	and
	\begin{align}
	\Omega_{R} = \frac{1}{4}\sqrt{(A - D)^{2} + 4BC}
	\end{align}
where $\Omega_{R}$ is the two-photon Rabi frequency. Full contrast in the resulting population oscillation between the two states is achieved when $\Lambda \rightarrow 1$ which occurs for $A = D$. Setting Eq. \ref{eq:A} equal to Eq. \ref{eq:D}, we then solve for the $\delta_\text{ac}$, the differential light shift of the two ground states. This defines $\delta_{ac}$ in Eq. \ref{eq:eom_stark_nice}.

The two-photon Rabi frequency $\Omega_{R}$ is found by multiplying Eq. \ref{eq:B} and \ref{eq:C}, $4\Omega_R^2=BC$,
\begin{equation} \label{eq:solution}
\begin{split}
    4\Omega_R^2 = \sum_{n,m=-N+1}^{N} \frac{\Omega_{m-1}\Omega_{m}\Omega_{n-1}\Omega_{n}}{(\Delta-m\Delta^\text{Cs}_\text{hfs})(\Delta-n\Delta^\text{Cs}_\text{hfs})} \\
    \times e^{-i(2(m-n)k_\text{hfs}z+\phi_m-\phi_{m-1}-\phi_n+\phi_{n-1})},
\end{split}
\end{equation}
where we have used that $2(m-n)k_\text{hfs}=k_m+k_{m-1}-k_n-k_{n-1}$. Using the symmetry of indices $m$ and $n$, the terms in the sum can be organized into conjugate pairs that are rewritten as cosines, and we arrive at the solution presented in Eqs. \ref{eq:eom_rabi_nice} and \ref{eq:Rmn}. This describes how each sideband pair affects the total two-photon Rabi frequency for intracavity Raman transitions.

%The Hamiltonian is written as
%\begin{equation}
%\hat{H}(t) =
%    \begin{pmatrix}
%E_1^0 & 0 & V_{13}(t) \\
%0 & E_2^0 & V_{23}(t) \\
%V_{13}(t) & V_{23}(t) & E_3^0
%\end{pmatrix}.
%\end{equation}

%\begin{align}
%    i\hbar \frac{d}{dt}C_i(t) =& V(t) C_3(t)e^{-i\omega_{13}t} \\
%    i\hbar \frac{d}{dt}C_i(t) =& \delta C_2(t) + V(t) %C_3(t)e^{-i\omega_{23}t} \\
%i\hbar \frac{d}{dt}C_3(t) =&V(t) C_1(t)e^{i\omega_{13}t} + V(t) %C_2(t)e^{i\omega_{23}t}
%\end{align}
	\begin{figure}[b]
	    \centering
	    \includegraphics[width=1\linewidth]{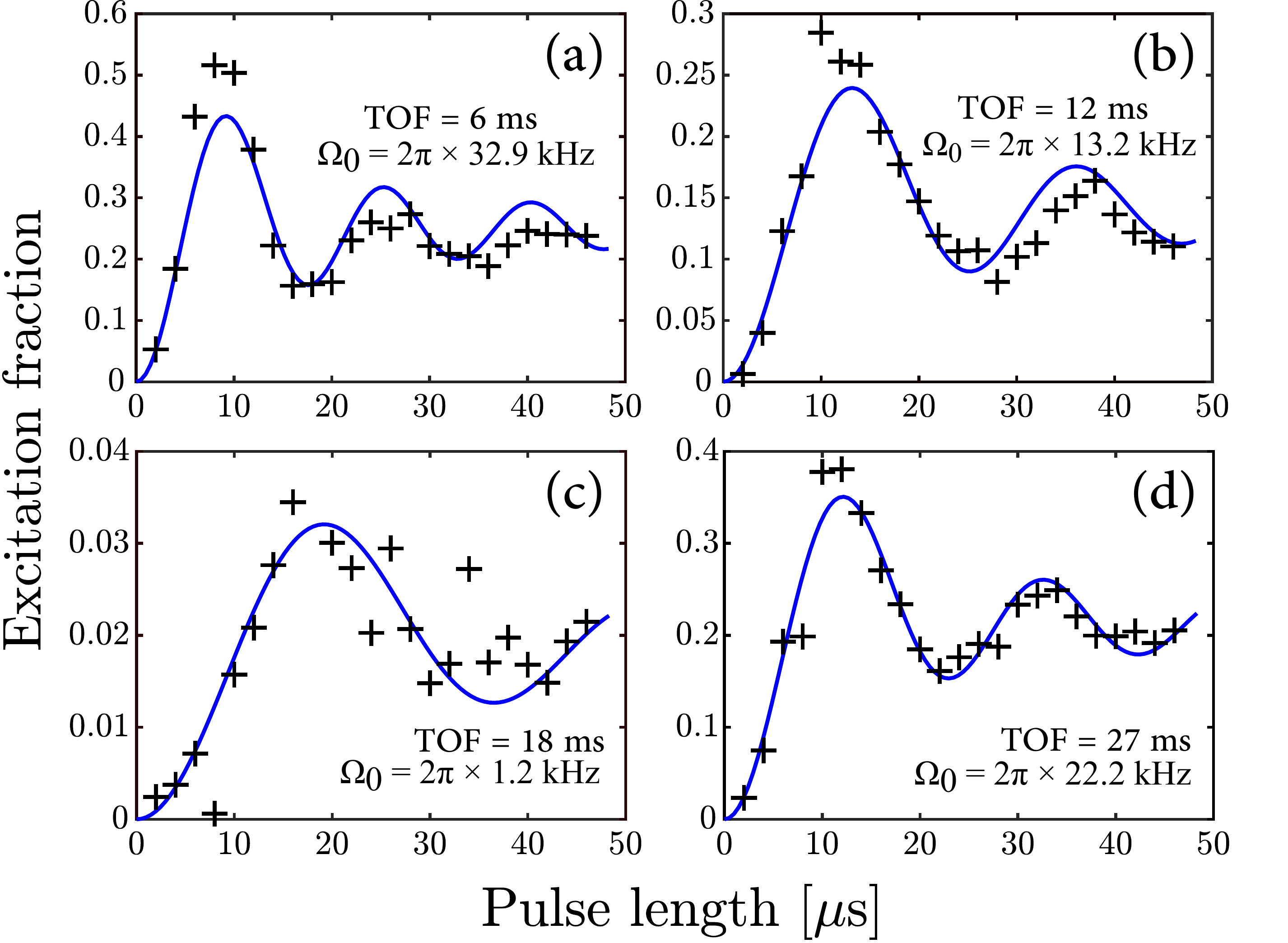}
	    \caption{Fitting the Rabi floppings for different time of flights (TOF). (a), (b) and (d) are at spatial positions with medium to high Rabi frequencies, while (c) is in the dead zone. For such a low Rabi frequency, decoherence across the atomic cloud leads to a very low peak excitation fraction (note the different y-scaling on the plots).}
	    \label{fig:fitexample}
	\end{figure}

\section{Rabi frequency fit model} \label{sec:fit_appendix}
The fit model accounts for the comparable sizes of the atom cloud and the Raman beams' waist. We model the atom cloud density as a spherical Gaussian, given by 
	\begin{equation} \label{eq:spherical_gaussian}
	n_{\text{atom}}(r) = \frac{N_{0}}{\pi^{3/2}\sigma^{3}} \exp\left( - \frac{r^{2}}{\sigma^{2}} \right)
	\end{equation}
	\noindent
	where $\sigma$ is the $\frac{1}{e}$ radius of the atom cloud, $N_{0}$ is the total number of atoms in the cloud, and $r$ is the position displacement from the center of the cloud. The Gaussian beam profile has intensity given by
	\begin{align}
	I(\rho) 	&= I_{0} \left( \frac{w_{0}}{w(z)} \right)^{2} \exp \left( -\frac{2\rho^{2}}{w(z)^{2}} \right) \nonumber \\
	&\approx I_{0} \exp \left( -\frac{2\rho^{2}}{w_{0}^2} \right)
	\end{align}
	\noindent
	where $\rho$ is the distance from the beam center transverse to the propagation axis. In the second line, we have assumed a weakly diverging beam such that $w(z) \approx w_{0}$ (in our experiment, the Rayleigh range of the cavity mode is $z_{R} = 1.90$ m, justifying this assumption). The waist of our cavity is $w_{0} = 718\text{ }\mu$m.
	The intensity $I_{\text{atom},i}$ seen by a given atom $i$ depends on its coordinate $\rho_{i}$, which we can parameterize by
	\begin{equation}
	I_{\text{atom},i} = \alpha_{i} I_{0},
	\end{equation}
	\noindent
	where $\alpha_{i} = \exp\left(-\frac{2\rho_{i}^{2}}{w_{0}^{2}}\right)$ is a unitless variable $\in [ 0, 1 ]$. Given the atom distribution Eq. \ref{eq:spherical_gaussian}, it can be shown \cite{JaffeThesis2018} that the probability distribution $f_{A}(\alpha)$ for the variable $\alpha$ across the atom cloud is
	\begin{equation}
	f_{A}(\alpha) = x \alpha^{x-1},
	\end{equation}
	\noindent
	where $x = \frac{w_{0}^{2}}{2\sigma^{2}}$.
	
	Each atom undergoes Rabi flopping according to 
	\begin{equation}
	P(\alpha, t) = \frac{\alpha \Omega_{R,0}}{\tilde{\Omega}_{R,\alpha}} \sin^{2}\left(\frac{1}{2} \tilde{\Omega}_{R,\alpha} t \right)
	\end{equation}
	\noindent
	where $\Omega_{R,0}$ is the two-photon Rabi frequency at the center of the beam, and $\tilde{\Omega}_{R,\alpha} = \sqrt{\left( \alpha \Omega_{R,0}\right)^{2} + \left( \alpha \delta_{\text{ac}} \right)^{2}}$ is the generalized two-photon Rabi frequency which includes an $\alpha$-dependent ac Stark shift. The average probability $P$ over the cloud can then be found by integrating over the $\alpha$ distribution,
	\begin{equation} \label{eq:Rabi_fitmodel}
	P(t) = \int_{0}^{1} \! d\alpha \, f_{A}(\alpha) P(\alpha, t).
	\end{equation}

	Included in our model is the finite, time-dependent size of the atom cloud. We assume a Gaussian distribution of velocities as well, such that the cloud size is given by
	\begin{equation}
	\sigma(t) = \sqrt{ \sigma_{0}^{2} + \sigma_{\text{v}}^{2}t^{2}},
	\end{equation}
	\noindent
	where $\sigma_{0} = 300\text{ }\mu$m is the initial size of the cloud at $t = 0$, and $\sigma_{\text{v}} = \sqrt{\frac{k_{\text{B}}T}{m_{\text{Cs}}}}$ is the velocity $\frac{1}{e}$ spread at temperature $T = 300$ nK, set by the temperature after Raman sideband cooling \cite{Kerman2000}. $k_{\text{B}}$ is the Boltzmann constant, and $m_{\text{Cs}}$ is the mass of the cesium-133 atom.
	
	We fit measured data to the model Eq. \ref{eq:Rabi_fitmodel} to extract the two-photon Rabi frequency $\Omega_{R,0}$ as a function of cavity parameters. $\Omega_{R,0}$ and $\delta_{\text{ac}}$ are the only fit parameters. In Fig. \ref{fig:fitexample} such fits are shown. For very low Rabi frequencies (Fig. \ref{fig:fitexample}c) the relative uncertainty on the fit increases, since the peak of the excitation fraction is very low. This makes it harder to distinguish between a slightly higher Rabi frequency or a detuning induced by the ac Stark shift. 
	
	\newpage

	\bibliographystyle{apsrev4-1}
	\bibliography{biblio_cavity_raman}
	
	%\vax{vicky's voice}
	%\slk{sofus' speculations}
	%\msj{matt's mumbling}
	%\hmu{holger's honing}
	%\cp{cris' critiques}
	%\todo{here's a thing to do}
	
	%- Sideband transitions are offset from zero
%- source mass add something

\end{document}